# The Antimatter Gravitational field

Piero Chiarelli

*National Council of Research of Italy, Area of Pisa, 56124 Pisa, Moruzzi 1, Italy*

*Interdepartmental Center "E. Piaggio" University of Pisa*

Phone: +39-050-315-2359

Fax : +39-050-315-2166

Email: pchiare@ifc.cnr.it.

**Abstract.** By using the gravity equation for quantum mechanical systems that takes into account the non-local interaction, the paper derives the characteristics of the antimatter in limit of macroscopic classical gravity.

The output of the theory shows that:

i. The antimatter has an attractive Newtonian potential;

ii. The antimatter attractive gravity is compatible with the CPT symmetry.

## 1. Introduction

If on one hand the general relativity has led to a satisfactory description of the gravitational interaction of macroscopic classical masses, on the other hand the quantum mechanics has brought to the physical definition of the antimatter states. Nevertheless, the leaking of a unified quantum gravity theory does not allow to clearky esthablish the characteristics of the matter-antimatter interaction.

One of the main problem of the quantum gravitational models is to produce theoretical outputs that lead to experimental evidence or confirmation [1-8].

It exists the objective difficulty of finding the physical ambit where the quantum mechanics and the gravitational effects are contemporaneously important. This happens because such theories describe a typology of events that own a quite different physical scale. One possibility, in order to find quantum-gravitational phenomena is to look at the microscopic dimensions where they become physically coupled such as the Planck scale.

By using the quantum gravitational equations (QGEs) obtained with the help of the hydrodynamic quantum formalism, the author showed [9] that the quantum effects play an important role in the collapse of a black hole since they oppose themselves to it by generating a repulsive force. This fact during the collapse of a mass of very small size (below the Planck mass) may hinder the formation of the black hole.

Another measurable output that can come from the quantum gravity is the detailed behavior of the gravitational field of antimatter. Many and discordant are the hypotheses on the gravitational features of the antimatter [10-14] and they cannot be resolved without a defined set of quantum gravitational equations. Actually, the proposed outputs are quite confusing since, if on one side, Cabbolet [5] claims that the CPT symmetry is incompatible with the matter-antimatter gravitational repulsion, on the other side, Villata [15-16] shows that the CPT agrees with anti-gravity .

Recently, by using the principle of minimum action, the author has derived from the hydrodynamic representation of quantum mechanics [17} the gravitational equation (QGE) that contains the non-local interactions of the quantum mechanics and gives an analytical connection between gravity and the fields of the matter [18]. The proposed theory leads to the compatibility of the enormous cosmological constant value deriving by the zero-point vacuum energy density of the QFT with respect the observed astronomical observations [19]. Since the QGE contains the explicit coupling with the particles fields, the characteristics of the antimetter gravity can analytically derived.

In this paper the author analytically derives the gravitational field generated by antimatter in order to give a contribution on the subject as well as to give physical outputs that can be experimentally verified.

## 2. The gravity of a scalar unchrged boson

The QGE, that includes the quantum potential energy for the definition of the space-time geometry, derived in ref. [17}, reads

$$R_{\sim€} - \frac{1}{2} R g_{\sim€} - \frac{8ƒ G}{c^4}\left(\overline{T}_{\sim€} + \Lambda g_{\sim€}\right) = 0 \qquad (2.0.1)$$

where

$$\overline{T}_{\sim€} = T_{\sim€} - \frac{1}{3} g_{\sim€} T_{S}^{S} = -|Œ|^2 \dot{q}_{\sim} p_{€} \;, \qquad (2.0.2)$$

$$T_{\sim€} = -|Œ|^2 \left( \dot{q}_{\sim} \frac{\partial L}{\partial \dot{q}^{€(k)}} - L g_{\sim€} \right) \qquad (2.0.3)$$

and where the *cosmological energy-impulse tensor density* (CEITD) $\Lambda g_{\sim€}$ reads

$$\begin{aligned} \Lambda g_{\sim€} &= \left(\overline{T}_{\sim€} - T_{\sim€}\right) + \frac{1}{3} T_{class\,S}^{\;S} g_{\sim€} \\ &= \frac{1}{3}\left(-T_{S}^{S} + T_{class\,S}^{\;S}\right) g_{\sim€} \end{aligned} \;. \qquad (2.0.4)$$

where

$$T_{class\,S}^{\;S} = lim_{\hbar \to 0} \overline{T}_{S}^{S} \;. \qquad (2.0.5)$$

In formula (2.0.3) $L = D_t S = -p_{\sim} \dot{q}_{€}\, g^{€\sim}$ where $S = \frac{i\hbar}{2} ln \frac{Œ}{Œ *}$ , where $\dot{q}^{˜}$ and $p_{\sim} = -\frac{\partial S}{\partial q^{˜}}$ are given by the solution of the quantum hydrodynamic problem [17] (i.e., $Œ = |Œ| exp \frac{i}{\hbar} S$ and $L = \frac{dS}{dt}$, whereŒ obeys to the Klein Gordon equation (KGE)

$$\left(\partial^{˜} Œ\right)_{;\sim} = \frac{1}{\sqrt{-g}} \partial_{\sim}\left(\sqrt{-g}\, g^{\sim€} \partial_{€} Œ\right) = -\frac{m^2 c^2}{\hbar^2} Œ \qquad (2.0.6)$$

that in the hydrodynamic representation[17-18], as a function of $|Œ|$ and $\partial_{€} S$ read [17]

$$g_{\sim€} \partial^{€} S\, \partial^{˜} S - \hbar^2 \frac{1}{|Œ| \sqrt{-g}} \partial_{\sim} \sqrt{-g}\left(g^{\sim€} \partial_{€} |Œ|\right) - m^2 c^2 = 0 \,. \qquad (2.0.7)$$

$$\frac{1}{\sqrt{-g}}\frac{\partial}{\partial q^{\mu}}\sqrt{-g}\left(g^{\mu\nu}|\psi|^{2}\partial_{\nu}S\right)=0 \qquad (2.0.8)$$

see below (2.0.8-9) )

For k-plane waves $\psi_{(k)}=|\psi_{(k)}|exp\left[ik_{\mu}q^{\mu}\right]$, (2.0.3-4) read, respectively,

$$T_{\mu\nu(k)}=-\frac{m|\psi_{k}|^{2}c^{2}}{\chi}\sqrt{1-\frac{V_{qu}}{mc^{2}}}\left(u_{\mu}u_{\nu}-g_{\mu\nu}\right) \qquad (2.0.9)$$

$$\Lambda_{(k)}=-\frac{mc^{2}}{\chi}|\psi_{k}|^{2}\left(1-\sqrt{1-\frac{V_{qu(k)}}{mc^{2}}}\right). \qquad (2.0.10)$$

where

$$V_{qu}=-\frac{\hbar^{2}}{m}\frac{1}{|\psi|\sqrt{-g}}\partial_{\mu}\sqrt{-g}\,g^{\mu\nu}\partial_{\nu}|\psi| \qquad (2.0.11)$$

## 3. The Galilean limit of the antimatter gravitational field

When we perfom the classical limit of quantum gravity we obtain the general relativity theory both for matter and for antimatter. Thence, the character of the gravity between (quantum de-coupled) matter and antimatter can be obtained by the Galilean limit of the classical gravitational equation.
Morever, since in the classical macroscopic dynamics, the quantum decoherence takes place, we consider only stable states (i.e., eigenstates) that can survive at the macroscopic level [20].
In the following we derive the characteristics of classical matter-antimatter gravitational field for a sclara uncharde particle (i.e., disregarding the interaction due to the charge and the spin of the particles. In this case the boson EITD (3) reads

$$T_{\mu\ \pm}^{\ \nu}=(\pm)-\frac{m|\psi_{\pm}|^{2}c^{2}}{\chi}\sqrt{1-\frac{V_{qu}}{mc^{2}}}\left(u_{\mu}u^{\nu}-\delta_{\mu}^{\ \nu}\right). \qquad (3\text{-}0.1)$$

where $\psi_{+}$ and $\psi_{-}$ are the eigen-functions of positive and negative energy states..
By using the conditions of the classical limit

$$\hbar\rightarrow 0 \text{ and } V_{qu}\rightarrow 0\ , \qquad (3\text{-}0.2)$$

and of the Galilean limit, that actually means low energy limit, it holds

$$\chi\cong 1\ . \qquad (3\text{-}0.3)$$

and

$$u_{\mu}=\left(u_{0},-u_{\alpha}\right)\sim\left(1,(0,0,0)\right), \qquad (3\text{-}0.4)$$

the QGEs for positive and negative energy state of particles reads [17], respectively,

$$\pm R_{\nu\mu}\mp\frac{1}{2}g_{\nu\mu}R_{\alpha}^{\ \alpha}\mp\frac{8\pi G}{c^{4}}\frac{m|\psi_{p\pm}|^{2}c^{2}}{\chi}g_{\mu\nu}=\frac{8\pi G}{c^{4}}T_{\nu\mu_{\pm}}=\pm\frac{8\pi G}{c^{4}}T_{\nu\mu} \qquad (3\text{-}0.5)$$

Since the antiparticle are described by particle negative energy states (see appendix) so that $Œ_p = Œ_+$ and vice versa $Œ_{ap} = Œ_-$ (3-0.5), in presence of a particle and its antiparticle (not quantum entangled*), equations lead to the overall QGE

$$R_{€~} - \frac{1}{2} g_{€~} R_{r}{}^{r} - \frac{8f\,G}{c^4} \frac{\left( m/Œ_+/^2 + m/Œ_-/^2 \right) c^2}{\text{X}} g_{~€} = \frac{8f\,G}{c^4} \left( T_{€~_+} - T_{€~_-} \right) = \frac{8f\,G}{c^4} T_{€~_{tot}} \qquad (3\text{-}0.8)$$

where the overall EITD is just the sum of each EITD [20] and, by using (3-0-1, 3-0.7), reads [17]

$$T_{€~_{tot}} = c^2 m \left( /Œ_+/^2 + /Œ_-/^2 \right) ( u_{~} u_{€} - g_{~€} ) \cong c^2 m \left( /Œ_+/^2 + /Œ_-/^2 \right), \qquad (3\text{-}0.9)$$

By introducing (3-0.9) in equation (3-0.9) it follows that

$$R_0{}^0 = \frac{4f\,G}{c^2} m \left( /Œ_+/^2 + /Œ_-/^2 \right) \left( 2u_0 u^0 - 1 \right) = \frac{4f\,G}{c^2} m \left( m/Œ_+/^2 + /Œ_-/^2 \right) \qquad (3\text{-}0.10)$$

where $G$ is the gravitational constant.
Moreover, given that the Galilean gravitational potential $\{$ , is a function of the component $g_{00}$ of the metric tensor [21] as follows

$$g_{00} = 1 + \frac{2\{}{c^2} \qquad (3\text{-}0.11)$$

whose trace, in the Galilean limit at zero order, can be approximated as

$$g_{rr} \cong -2, \qquad (3\text{-}0.12)$$

it follows that the QGE (21) reduces to

$$R_0{}^0 = -\frac{4f\,G}{c^2} m \left( /Œ_+/^2 + /Œ_-/^2 \right) = R_{00} = \frac{\partial \Gamma^{r}{}_{00}}{\partial q^{r}} \approx -\frac{1}{2} \frac{\partial g_{\text{xx}} \frac{\partial g_{00}}{\partial q^{r}}}{\partial q^{r}} = \frac{1}{c^2} \frac{\partial}{\partial q^{r}} \frac{\partial \text{W}}{\partial q^{r}} \qquad (3\text{-}0.13)$$

that leads to

$$\frac{\partial}{\partial q^{r}} \frac{\partial \text{W}}{\partial q^{r}} = -4f\,Gm \left( /Œ_{p+}/^2 - /Œ_{ap+}/^2 \right). \qquad (3\text{-}0.14)$$

If we consider the case of point-like particle and antiparticle located in $R_+$ and $R_-$, respectively, with spatial densities

$$/Œ_+/^2 = \text{u}( r - R_p ) \qquad (3\text{-}0.15)$$

$$/Œ_-/^2 = \text{u}( r - R_{ap} ), \qquad (3\text{-}0.16)$$

it follows that

$$\frac{\partial}{\partial q^{r}}\frac{\partial W}{\partial q^{r}} = -4f\,Gm\Big(u(\,r-R_p\,)+u(\,r-R_{ap}\,)\Big) \qquad (3\text{-}0.17)$$

and, by integration, that

$$W = -Gm\left(\frac{1}{R_p}+\frac{1}{R_{ap}}\right). \qquad (3\text{-}0.18)$$

where the gravitational field (3-0.18) generated just by the particle or the antiparticle reads, respectively,

$$W_p = -G\frac{m}{R_p} \qquad (3\text{-}0.19)$$

$$W_{ap} = -G\frac{m}{R_{ap}} \qquad (3\text{-}0.20)$$

that show an attractive gravitational potential for the antiparticle.
Moreover, calculating the radial force $F_r$ generated between two point-like masses of matter and antimatter (??-??), by posing and $R = /R_p - R_{ap}/$, it follows that

$$F_r = -\frac{\partial U}{\partial R} = -\frac{\partial \sum_i \frac{1}{2}\int m_i W_i\, dV}{\partial R} = -\frac{\partial \frac{1}{2}\left(\int m W_{ap}\, dV + \int m W_p\, dV\right)}{\partial R}$$

$$= -\frac{\partial \frac{1}{2}\left(m\int u(\,r-R_p\,) W_{ap}\, dV + m\int u(\,r-R_{ap}\,) W_p\, dV\right)}{\partial R}$$

$$= -\frac{\partial \frac{1}{2} m\left(\int u(\,r-R_p\,) W_{ap}\, dV + \int u(\,r-R_{ap}\,) W_p\, dV\right)}{\partial R} \qquad (3\text{-}0.21)$$

$$= \frac{\partial \frac{G}{2} m^2\left(\int u(\,r-R_p\,)\frac{1}{/\,r-R_{ap}\,/}dV + \int u(\,r-R_{ap}\,)\frac{1}{/\,r-R_p\,/}dV\right)}{\partial R}$$

$$= \frac{\partial \frac{G}{2} m^2\left(\frac{1}{/\,R_p-R_{ap}\,/}+\frac{1}{/\,R_{ap}-R_p\,/}\right)}{\partial R} = -\frac{Gm^2}{R^2}$$

where $U$ is the gravitaional potential.

## Conclusion

In the present work the gravitational field of the antimatter in the Galilean limit is derived by the quantum gravity equations. The results show that time inversion bringing a scalar boson into its antiparticle, is compatible with an attractive gravitational potential. Since for charged particles with spin the CPT transforms a particle in its antiparticle, the paper shows that the matter-antimatter attractive gravitational potential can be compatible with the CPT symmetry.